\documentclass{article}
\usepackage{iclr2027_conference,times}
\usepackage[T1]{fontenc}
\usepackage{amsmath,amssymb,booktabs,tabularx,longtable,array}
\usepackage{graphicx,microtype,float}
\usepackage[hidelinks]{hyperref}
\usepackage{url}
\hypersetup{
  pdftitle={The Last Mile Is the File: OfficeEditBench for Preservation-Aware Office Editing},
  pdfauthor={Zhiwen Wu and Chengxu Wu}
}
\newcommand{\bench}{OfficeEditBench}
\newcommand{\ind}{\mathbf{1}}
\newcolumntype{L}{>{\raggedright\arraybackslash}X}

\title{The Last Mile Is the File: OfficeEditBench\\for Preservation-Aware Office Editing}
\author{Zhiwen Wu\thanks{Equal contribution.}\\
Peking University\\
\texttt{2501210427@stu.pku.edu.cn}
\And
Chengxu Wu\footnotemark[1]\\
Peking University\\
\texttt{2501210417@stu.pku.edu.cn}}
\iclrfinalcopy
\begin{document}
\maketitle
\begin{abstract}
A small Office edit creates two obligations: propagate every required update and leave protected state untouched. Updating too little leaves dependencies inconsistent; updating too much changes content the user did not authorize. We introduce \bench{}, a 170-task benchmark for \emph{change-scoped maintenance} of spreadsheets, presentations, and documents. Task contracts specify required updates, protected state, native structures, and applicable interaction requirements. Across 510 archived task--system outcomes from WorkBuddy, Doubao, and Codex, we distinguish file delivery, target completion, and verifier-defined acceptance. Hard package-valid delivery ranges from 92\% to 100\%, yet no selected output satisfies the complete contract. Case analysis highlights why local correctness is insufficient: an updated value can lose its generating formula, a revised rule can fail to reach related conclusions, and a new deadline can omit a retained prerequisite. These mechanisms connect artifact-level checks to the continued maintainability of Office files. We analyze maintenance failures while distinguishing frozen automatic verdicts from human acceptability. \bench{} provides a testbed for completing required changes while preserving the logic and scope of existing work.
\end{abstract}

\section{Introduction}
\begin{figure}[t]
\centering
\includegraphics[width=\textwidth]{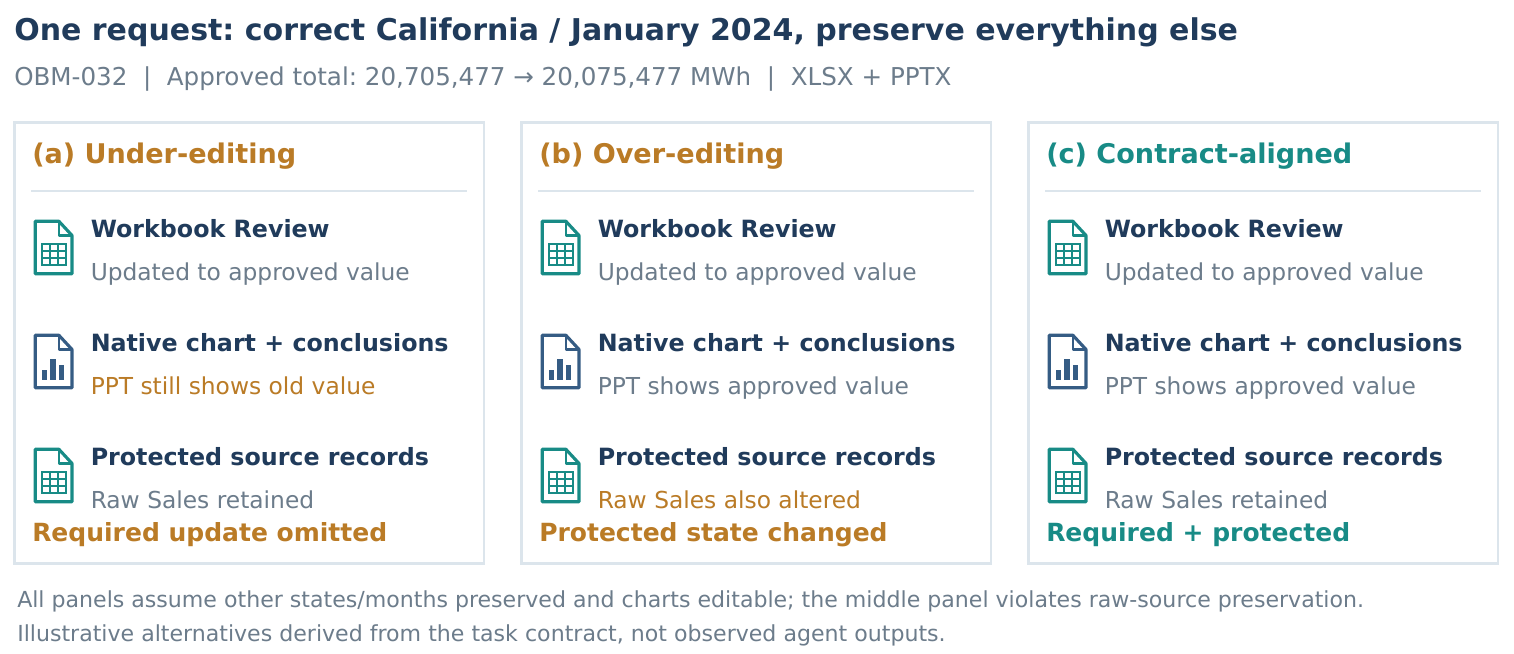}
\caption{\textbf{One request, two ways to miss its scope (OBM-032).} The approved value must propagate to the workbook, native PPT chart, and conclusions. Updating only the workbook omits required changes; also changing raw Sales violates preservation. The right panel satisfies the depicted obligations, assuming other protected state and native editability are retained. These are task-derived alternatives, not actual agent outputs, Office screenshots, or an empirical failure-frequency claim. The task requires XLSX and PPTX only.}
\label{fig:case}
\end{figure}

A business analyst asks an agent to correct California's January electricity total in a review workbook and update the meeting slides. The authoritative records imply 20,075,477 MWh, while the workbook reports 20,705,477. The approved correction must reach the workbook, the presentation's native chart, and related conclusions; raw source records, other states, and other months must remain unchanged. Updating only the workbook is \emph{under-editing}; updating those targets but also rewriting raw records is \emph{over-editing}. Figure~\ref{fig:case} illustrates these alternatives for benchmark task OBM-032, without attributing them to observed agent runs. Neither satisfies the full request.

Existing Office files contain prior work that the new request does not supersede. Formulas, chart data, notes, table structure, and approvals can remain valid even when one value needs correction. The objective is not merely to minimize edits: necessary changes may span multiple files. It is to \emph{complete every required update within its authorized scope while preserving protected content and native editability}. We call this \emph{change-scoped maintenance}. Required consequences, permitted changes, and protected state are distinct; authorization alone does not make every possible edit mandatory.

Execution-based benchmarks already evaluate changes to persistent environments. SWE-bench checks issue resolution and retained software behavior \citep{jimenez2024swebench}; spreadsheet and office benchmarks evaluate realistic manipulation and application workflows \citep{ma2024spreadsheetbench,wang2024officebench}. We build on this maintenance perspective rather than claim that preservation is a new evaluation principle. Our focus is its concrete realization across heterogeneous, editable Office artifacts: task-specific target regions, specified cross-representation consequences, protected observations, and delivery or clarification requirements.

\bench{} contains 170 tasks across three tiers and native XLSX, PPTX, and DOCX formats. It pairs strict task acceptance with decomposed check outcomes so that a readable package, a correct target, and a complete edit can be distinguished. The benchmark does not prescribe an editing algorithm; it evaluates the files and required interactions named by each contract.

We investigate three questions: \textbf{RQ1}, does file delivery imply complete maintenance? \textbf{RQ2}, what failures remain after target completion? \textbf{RQ3}, which maintenance obligations explain the gap? The central comparisons are between criteria applied to the \emph{same selected outputs}, complemented by case-level analysis of formulas, shared definitions, and retained conditions.

\paragraph{Contributions.}
We contribute (i) a 170-task suite that instantiates change-scoped maintenance in native Office workspaces; (ii) an evaluation design separating required updates, protected state, and task-level acceptance; and (iii) an analysis of 510 archived task--system outcomes that links aggregate results to specific maintenance obligations. The accompanying package preserves frozen verdicts, reproducible aggregates, provenance, and focused case evidence.

\paragraph{Open-source release.}
The benchmark, verifiers, and release documentation are available from the public repository:
\begin{center}
\href{https://github.com/Aniriswu/OfficeEditBench}{\texttt{https://github.com/Aniriswu/OfficeEditBench}}
\end{center}

\section{Related work}
\paragraph{Maintaining an existing state.}
SWE-bench evaluates repository edits using both fail-to-pass and pass-to-pass tests \citep{jimenez2024swebench}. Its regression-testing principle is directly relevant: satisfying a new requirement does not authorize breaking existing behavior. \bench{} instantiates this principle for native Office objects and task-specific business evidence. The challenge is representational as well as semantic: a displayed value can be correct even when its formula, chart linkage, or protected context is not preserved.

\paragraph{Office and interactive work.}
SpreadsheetBench and SpreadsheetBench~2 evaluate realistic spreadsheet manipulation and business workflows; OfficeBench studies cross-application automation \citep{ma2024spreadsheetbench,zhu2026spreadsheetbench2,wang2024officebench}. PPTArena studies in-place slide editing, PPT-Eval penalizes unnecessary modifications, and DeckEdit-Bench explicitly measures object preservation \citep{ofengenden2026pptarena,gandhi2026ppteval,kim2026editppt}. OmegaUse-OfficeVal combines multi-format deliverables, code-based checks, unintended-damage penalties, and expert--code calibration \citep{zhou2026omegauseofficeval}. Thus neither Office editing nor preservation checking is new in isolation. Our focus is \emph{change-scoped maintenance}: task-specific authorized consequences and protected observations, with strict conjunctive acceptance alongside partial diagnostics. Appendix~\ref{app:related} compares these overlapping designs. WebArena, OSWorld, and WorkArena++ provide broader execution-based settings \citep{zhou2024webarena,xie2024osworld,boisvert2024workarena}.

\paragraph{Protocols and evaluation validity.}
$\tau$-bench evaluates interactive tool use through database end states and repeated-trial reliability \citep{yao2025taubench}. Our specified clarification tasks are narrower: they use fixed operator replies, and the archived snapshots do not estimate repeated-trial reliability. MLE-bench separates the submission interface from the agent implementation and explicitly documents execution conditions \citep{chan2025mlebench}. ScienceAgentBench uses expert validation and human assessment alongside automatic measures \citep{chen2025scienceagentbench}. These works highlight distinct requirements for a useful benchmark: an operational task definition, interpretable outcomes, and evidence that the measurement is valid. We provide task-level diagnostics and construction checks while explicitly identifying the remaining validation gap. Appendix~\ref{app:related} compares evaluation units without treating results on different tasks as comparable baselines.

\section{Benchmark design and construction}
\label{sec:design}
\begin{figure}[t]
\centering
\includegraphics[width=\textwidth]{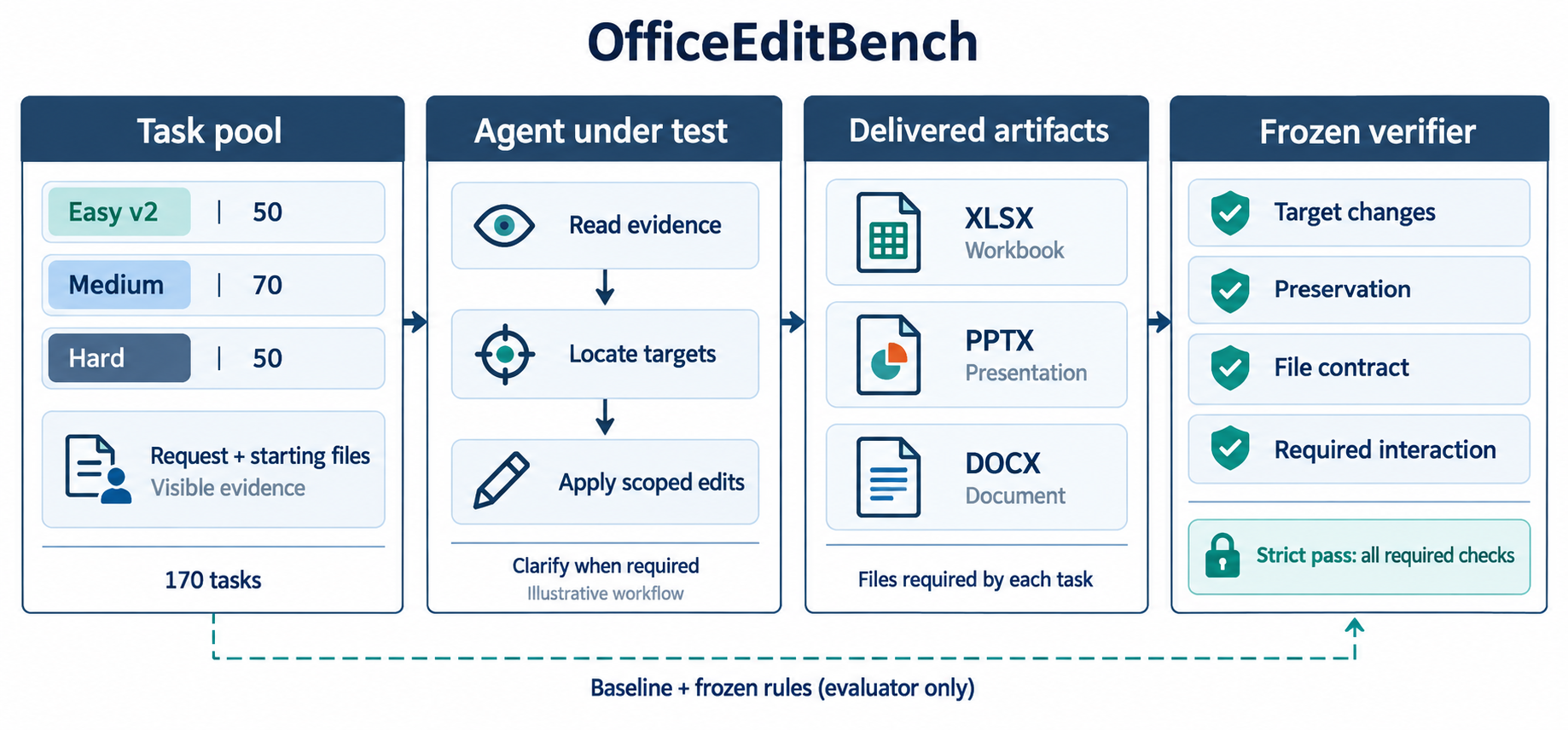}
\caption{\textbf{\bench{} evaluation architecture.} Requests, starting files, and visible evidence define the agent input; submitted native artifacts are checked against the task contract. Strict success requires all applicable predicates, including interaction when specified. Agent steps are illustrative. The dashed path marks intended evaluator-only access, not demonstrated historical isolation.}
\label{fig:architecture}
\end{figure}

\subsection{Task contract and design principles}
A task is $T=(W_0,R,E,A,V)$: materialized starting artifacts $W_0$, request $R$, visible evidence $E$, authorization specification $A$, and versioned verifier $V$. The agent returns edited artifacts $\widehat W$ and, when required, an interaction record (Figure~\ref{fig:architecture}). Three principles organize the contract.

\paragraph{Authorized change includes specified consequences.}
The target region comprises directly requested objects and their authorized dependents. For OBM-032, the workbook value, native presentation chart, and related conclusions must agree with the approved evidence (Figure~\ref{fig:case}). Agreement alone is insufficient if all locations repeat the same wrong value. Dependencies are encoded in instructions, evidence, and predicates; the benchmark does not infer a complete dependency graph from arbitrary files.

\paragraph{Preservation is relative to the actual starting state.}
Let $\mathcal T_T$ denote target predicates and $\mathcal Q_T$ protected observations. The artifact obligations are
\begin{equation}
 \forall t\in\mathcal T_T:\ t(\widehat W;R,E,A)=1,
 \qquad
 \forall q\in\mathcal Q_T:\ q(\widehat W)\equiv_T q(W_0).
\end{equation}
The task-specific relation $\equiv_T$ is the implemented comparison, not universal semantic equivalence or byte identity. Protected state can include non-target values, formulas, text, notes, and native structures. Reference comparisons use materialized task inputs so that fixture preparation is not mistaken for an agent edit.

\paragraph{The deliverable is an editable artifact.}
The task names required files and any interaction obligations. File validity, target correctness, and preservation are distinct checks: a parseable Office package does not by itself establish a completed edit or native-application usability. When clarification is required, the specified concepts and fixed reply are checked; this is not full verification of every action in a trajectory.

\subsection{Coverage and construction}
Table~\ref{tab:tiers} summarizes the active suite. Easy v2 replaces the retired easy set. Its 50 tasks contain 25 English and 25 Chinese instructions with localized edits to dates, values, formulas, slide content, native tables, or document headings and action lines. Medium comprises a 10-task pilot and three 20-task cohorts: operational workbooks, cross-media updates, and sector reports. Medium and Hard instructions are predominantly Chinese.

\begin{table}[t]
\centering\small
\caption{\textbf{Active task inventory.} Tiers express construction intent, not a calibrated difficulty scale. Format denotes the primary artifact for Medium; some tasks include linked pairs.}
\label{tab:tiers}
\begin{tabularx}{\textwidth}{p{.13\textwidth}rXX}
\toprule
Tier & Tasks & Artifact coverage & Contract emphasis\\
\midrule
Easy v2 & 50 & 20 XLSX, 20 PPTX, 10 DOCX; one file & Local target/type correctness; non-target content and structure\\
Medium & 70 & 30 spreadsheet-, 20 presentation-, 20 document-primary & Contextual business edits; linked native objects and protected narrative\\
Hard & 50 & XLSX + PPTX + DOCX per task; 2--3 evidence files & Cross-file consequences, approval boundaries, source authority\\
\bottomrule
\end{tabularx}
\end{table}

Hard combines four shared source scenarios---public-company reporting (13 tasks), retail operations (13), capital/procurement (12), and environmental compliance (12)---with fourteen workflow profiles. Profiles cover dependent and independent changes, partial approvals, late amendments, evidence repair, release decisions, and duplicate labels. Eight tasks require first-turn clarification. The 50 task IDs are variations over shared workspaces, not 50 independent organizations. File size is contextual workload, not our definition of edit difficulty.

\paragraph{From source material to a checked task.}
The repository organizes task construction around source fixtures, task-specific starting states, requests and authority evidence, and target/protected-state predicates. For Hard, SEC, UCI retail, Chicago procurement, and EPA-derived material provides source context; approvals, personnel, and version chains are synthetic workflow overlays. Materialization adapts shared fixtures to each request. The task then specifies which evidence is authoritative, what must change, what must remain unchanged, and which files must be returned. Reference outputs and negative probes exercise the resulting predicates where archived QA is available (Section~\ref{sec:qa}). This is a description of the implemented components, not a claim of independent annotation or uniform QA coverage across all tiers. Appendix~\ref{app:dataset} records source boundaries and profile coverage.

\paragraph{Request provenance.}
Requests are benchmark-authored around public-data-derived or synthetic workspaces. Public sources ground selected business contexts; we do not claim a sampled distribution of naturally occurring user requests. Source boundaries and construction records are documented in Appendix~\ref{app:dataset}.

\section{Evaluation protocol and construction checks}
\subsection{Strict acceptance and diagnostic scores}
For required predicates $\mathcal K_T$ and interaction record $H$, strict acceptance is
\begin{equation}
 A_T(\widehat W,H)=\prod_{j\in\mathcal K_T}\ind[j(\widehat W,H)\text{ passes}].
\end{equation}
We report the fraction of selected outputs with $A_T=1$ as the \emph{strict pass rate}: throughout, this means \emph{verifier-defined acceptance}. Partial scores summarize which requirements were satisfied; they do not replace the conjunction. This selected-output statistic is neither pass@1 nor a human-acceptability rate.

Easy v2 assigns 10 points to readability, 10 to structure, 70 to the target, and 10 to non-target preservation. Target checks include required types and formatting. A failed PPTX structural guard records downstream target and non-target checks as failed because object correspondence cannot be established. Their substantive content correctness is \emph{unknown}, not independently observed to be wrong; we retain the stored scoring policy without interpreting these entries as separate content errors.

Medium and Hard use semantic ($S$) and preservation ($P$) categories. For category $\mathcal G$ with fixed weights $w_j$,
\begin{equation}
 G=100\frac{\sum_{j\in\mathcal G}w_j\ind[j\text{ passes}]}{\sum_{j\in\mathcal G}w_j}.
\end{equation}
Hard verifier \texttt{0.5.2-frontier50} uses $C=0.8S+0.2P$, or $C=0.7S+0.2P+0.1I$ on clarification tasks, where $I$ is the 0--100 interaction score. Medium uses cohort-specific predicates with pilot scores normalized to 0--100. Means are equal-task averages of stored scores. Different weights and requirements preclude interpreting tier composites as a common calibrated scale.

\subsection{Construction checks and their scope}
\label{sec:qa}
Archived Easy v2 QA accepts 50/50 reference outputs, rejects all 50 unchanged inputs, and rejects all 50 variants with a correct target plus an unrelated mutation. The last group averages 90/100 (Figure~\ref{fig:motivation}). Two additional probes reject a date-looking string and a literal replacing a required formula. Archived Hard QA accepts 50/50 references, rejects all 50 unchanged inputs ($S=0,P=100$), and rejects 12/12 adversarial variants spanning location, authorization, preservation, interaction, and missing files.

These are implementation checks, not independent estimates of evaluator validity. References and predicates may share assumptions; exact text, style identifiers, or formula representations may reject legitimate alternatives. No completed blind adjudication or comprehensive native-application rendering audit is available for the agent outputs. We retain the frozen verdicts, distinguish observed check failures from inferred causes, and specify an independent validation protocol in Appendix~\ref{app:measurement}.

\section{Evaluation results}
\label{sec:results}
\subsection{Evaluation scope and evidence}
The September~26, 2026 freeze comprises 170 tasks per system and 510 task--system outcomes. The three products are WorkBuddy, Doubao, and Codex; archived labels include \texttt{gpt-6-sol} for Codex and version 5.3.14 for WorkBuddy Medium. These are product snapshots, not matched underlying-model configurations.

\begin{figure}[t]
\centering
\includegraphics[width=\textwidth]{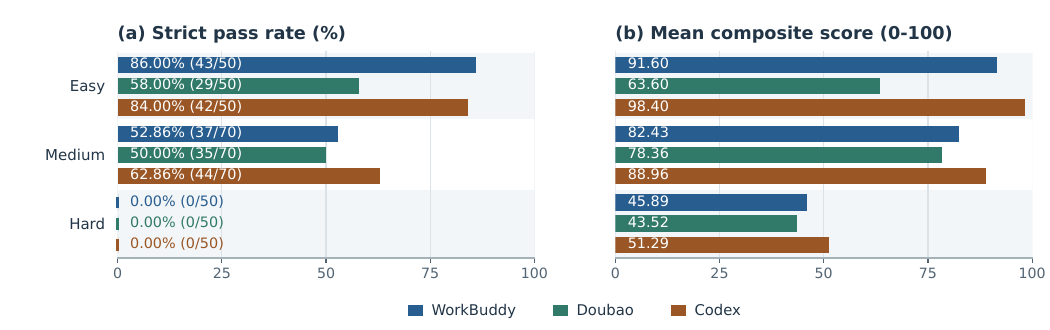}
\caption{\textbf{Three systems, three task tiers: strict acceptance and partial completion.} (a) Selected-output strict pass rates with pass/task counts. (b) Mean composite scores over every task in each tier, including non-delivery zeros. All 510 frozen outcomes are represented. Partial scores are tier-specific diagnostics, not a calibrated difficulty scale. WorkBuddy Hard is exploratory and Doubao Hard uses the corrected snapshot; submission-selection details are in Appendix~\ref{app:evidence}.}
\label{fig:results}
\end{figure}

We report selected-output results from all three agents on the same 170-task inventory under the frozen, tier-specific scoring criteria. Recorded retries, corrections, and submission-selection policies are retained in Appendix~\ref{app:evidence}; they do not constitute a uniform first-attempt protocol. The archive does not establish budget matching or uniform gold/verifier isolation, so these results are diagnostic rather than a controlled blind comparison. WorkBuddy Hard retains its exploratory designation, and Doubao Hard uses the corrected final snapshot. Figure~\ref{fig:results} summarizes strict acceptance and partial scores; Table~\ref{tab:results} adds component means and retains scored non-deliveries.

\begin{table}[t]
\centering\small
\caption{\textbf{Verifier-defined acceptance and partial scores.} Means are on a 0--100 scale within each tier; brackets give each component's observed denominator. Product order is fixed across tiers.}
\label{tab:results}
\begin{tabular}{llrrrrr}
\toprule
Tier & System & Strict pass & Rate (\%) & Mean $C$ & $S$ [$n$] & $P$ [$n$]\\
\midrule
Easy & WorkBuddy$^\dagger$ & 43/50 & 86.00 & 91.60 & -- & -- \\
Easy & Doubao & 29/50 & 58.00 & 63.60 & -- & -- \\
Easy & Codex & 42/50 & 84.00 & 98.40 & -- & -- \\
\midrule
Medium & WorkBuddy$^\dagger$ & 37/70 & 52.86 & 82.43 & 85.97 [70] & 71.81 [70] \\
Medium & Doubao & 35/70 & 50.00 & 78.36 & -- & -- \\
Medium & Codex & 44/70 & 62.86 & 88.96 & 88.08 [70] & 91.63 [70] \\
\midrule
Hard & WorkBuddy$^\dagger$ & 0/50 & 0.00 & 45.89 & 41.30 [46] & 82.75 [46] \\
Hard & Doubao$^\ddagger$ & 0/50 & 0.00 & 43.52 & 35.11 [50] & 73.17 [50] \\
Hard & Codex & 0/50 & 0.00 & 51.29 & 43.78 [50] & 79.84 [50] \\
 
\bottomrule
\end{tabular}
\par\smallskip
\begin{minipage}{\textwidth}\footnotesize
$\dagger$ WorkBuddy: operator-supplied atomic score records; Hard is exploratory. $\ddagger$ Doubao Hard: corrected final snapshot. Hard $C$ uses all 50 tasks; WorkBuddy Hard $S/P$ use 46 deliveries, whereas Doubao and Codex $S/P$ use all 50, including Codex's missing-output zero. Other reported means use the full tier. Dashes denote unavailable or non-comparable categories, not zero. Scores are not calibrated across tiers.
\end{minipage}
\end{table}

\subsection{RQ1: Does delivery imply complete maintenance?}
Hard provides the clearest separation: WorkBuddy, Doubao, and Codex deliver 46/50, 50/50, and 49/50 package-valid file triples, yet each records 0/50 strict passes (Figure~\ref{fig:gap}). Codex OBW4-045 fails all file-open checks and receives $C=S=P=0$; WorkBuddy's four non-deliveries contribute zero to $C$ but have unavailable component scores. Delivery accounting therefore matters both for strict outcomes and for the denominator of component means.

Semantic predicates remain unsatisfied even before preservation is considered. Neither Doubao nor Codex has a Hard task with $S=100$; their all-task semantic means are 35.11 and 43.78. WorkBuddy likewise has no $S=100$ result among its 46 scored deliveries. Thus, the zero-pass result cannot be attributed solely to preservation checks. Equally, it does not establish that every delivered artifact is useless: a strict conjunction can collapse substantially different partial outcomes to the same zero.

Preservation diagnostics locate additional mismatches. Among package-valid triples, non-target document-text predicates fail in 34/46 WorkBuddy, 45/50 Doubao, and 49/49 Codex cases; non-target spreadsheet predicates fail in 33/46, 44/50, and 35/49. These are non-exclusive task-level verdict counts, not counts of damaged objects or their severity. Appendix~\ref{app:cohorts} retains the predicate-specific denominators; Appendix~\ref{app:source-sensitivity} reports source-group results and descriptive leave-one-source-out sensitivity.

\subsection{RQ2: What remains after target completion?}
On Easy, Codex passes all 50 target checks but only 42 strict contracts. All eight failures are XLSX tasks scoring 90. Comparison of their input and output files identifies the same structural mismatch: the \texttt{Lists} worksheet changes from hidden to visible, while the other checked structural properties remain unchanged. This is a concrete state change beyond the requested cell edits (Appendix~\ref{app:recovered}), not evidence by itself that the workbook is unusable.

Medium provides a complementary view. WorkBuddy has 45 outputs with $S=100$, of which eight fail strict success; Codex has 49, of which five fail (Figure~\ref{fig:targetgap}). The corresponding conditional disagreement fractions are 8/45 (17.8\%) and 5/49 (10.2\%). These denominators comprise target-perfect outputs, not all 70 tasks, and do not support a controlled comparison between products. Comparable Doubao component data are unavailable. The WorkBuddy report's OBM-001 illustrates the distinction with $S=100$, $P=40$, and $C=85$; Codex's five cases are OBM-016, 029, 041, 047, and 050.

These disagreements motivate joint reporting of target and preservation predicates. WorkBuddy's operator attributes five Easy DOCX failures to style-ID renumbering. The atomic checks confirm non-target failures in all five and target-check failures in three, but do not store the identifiers or a rendered comparison. We therefore retain the frozen verdicts without treating the operator's explanation as an independently established root cause or equating these failures with visible layout damage.

\subsection{RQ3: Which maintenance obligations explain the gap?}
The WorkBuddy records distinguish successful scoped synchronization from three ways a locally correct edit can leave an incomplete deliverable (Table~\ref{tab:maintenance-cases}). Atomic checks support formula-behavior failures and record exact expected/actual text for the two document cases; the operator's analysis supplies additional editing context. Scores match the frozen records. These are selected diagnoses based on check records, not mechanism-frequency estimates or independent Office-file inspection.

\begin{table}[t]
\centering\small
\caption{\textbf{From local correctness to maintainable files.} WorkBuddy Medium cases combine atomic-check records with operator case descriptions. $C$ is the unchanged frozen composite, not a new human rating.}
\label{tab:maintenance-cases}
\begin{tabularx}{\textwidth}{p{.12\textwidth}rLL}
\toprule
Task & $C$ & Reported outcome & Maintenance obligation\\
\midrule
OBM-032 & 100 & Approved total synchronized across workbook, conclusions, native chart, and embedded data & Complete the authorized chain while retaining other states, months, and source records.\\
OBM-034 & 76.92 & Current values correct; operator reports replacing \texttt{AVERAGE} with a constant & Preserve the relationship to source data, not only today's displayed value.\\
OBM-058 & 62.50 & Rule says delay $\geq15$ minutes; a 15-minute flight still labeled on time in summary/timeline & Propagate a changed definition to its dependent statements.\\
OBM-062 & 0 & Deadline changed to 30 days after acceptance; compliant-invoice prerequisite omitted & Update the requested term without deleting a retained condition.\\
\bottomrule
\end{tabularx}
\end{table}

\paragraph{Preserve computation, not just the current answer.}
OBM-034 passes numerical recalculation and chart checks but fails formula-editability and source-response checks. The operator identifies replacement of an \texttt{AVERAGE} formula with a constant as the editing mechanism. The displayed result can be right at delivery while the workbook no longer responds to later source changes. This motivates checking formula behavior separately from numeric equality. In contrast, OBM-032 passes every stored check, showing a successful bounded cross-file synchronization under the task contract.

\paragraph{Propagate rules and retain prerequisites.}
OBM-058 passes the revised delay-threshold check, but the stored actual text still labels the 15-minute event as on time in paragraph 4 and table 0, cell (1,2). OBM-062's actual text in paragraph 14 and table 0, cell (1,1) gives the new 30-day term but omits the compliant-invoice prerequisite. The former misses a required consequence; the latter removes a retained condition. These specific textual differences instantiate the two sides of change-scoped maintenance without equating formatting-signature changes with business-content damage.

\paragraph{Target and preservation gaps overlap.}
The frozen WorkBuddy Medium records partition the 33 rejected tasks into eight with $S=100,P<100$, eleven with $P=100,S<100$, and fourteen with both components below 100. Thus 25 have semantic-check gaps and 22 have preservation-check gaps. This decomposition rules out a single-axis account of the failures, but the categories include structural predicates as well as content checks. The sector-report cohort's 5/20 strict result cannot be attributed entirely to business reasoning because reported style-ID changes also affect its preservation scores.

\paragraph{A partial business chain is not a complete package.}
In the exploratory WorkBuddy Hard records, OBW4-050 scores 90 and passes the approved France Actual 43,600 workbook/chart checks, while a review-note predicate and non-target predicates fail. OBW4-002 scores 61.82 and passes the approved 2,585 value checks but fails source-use and governance-record predicates. These are check-level mismatches, not findings that every rejected note is substantively wrong. OBW4-006 records a missing submission and scores zero; an operator-reported internal edit does not establish a delivered artifact's correctness. Appendix~\ref{app:workbuddy-cases} separates log-supported observations from additional operator context.

Strict acceptance and partial scores therefore answer different questions. On Easy XLSX, WorkBuddy and Codex both average 96 but pass 19/20 and 12/20; on Hard, all strict outcomes are zero while partial scores remain heterogeneous. We retain both views (Figure~\ref{fig:gap}); format and cohort slices appear in Appendix~\ref{app:cohorts}.

\section{Discussion and limitations}
\label{sec:limitations}
\paragraph{Implications for agent and evaluator design.}
The observations suggest treating edit scope as a first-class constraint: identify authoritative evidence, propagate specified consequences, and inspect protected state after editing. They also support reporting an acceptance vector alongside a scalar score. These are design implications from observed disagreements, not experimentally validated improvements to an agent. The benchmark's practical value is to make such hypotheses testable on native artifacts.

\paragraph{Measurement validity.}
A check failure is not automatically a business-content failure. Artifact inspection found a concrete precision-sensitive rejection in Codex OBM-037 (the exact approved value compared against a rounded reference) and wording sensitivity in OBM-035 (an exclusion expressed without the required literal cue). Appendix~\ref{app:recovered} preserves their values, predicates, and hashes. Frozen scores remain unchanged; these cases are not treated as evidence of omitted updates. Independent human/native-application calibration has not been conducted. Any subsequent checker revision should be versioned and applied uniformly.

\paragraph{Execution and population limits.}
Complete per-attempt budget and submission-selection records are unavailable across systems. Nominally shared settings alone do not establish matched realized budgets or a uniform attempt protocol; the results do not estimate standardized first-attempt or repeated-trial reliability. WorkBuddy Hard retains its exploratory designation. Hard reuses four source scenarios; tiers and languages are not controlled experimental factors. We report descriptive counts without causal rankings or independence-based significance claims. Source-family holdouts and fully logged executions are needed to measure generalization and reliability.

\paragraph{Exposure and release scope.}
Public gold enables reproduction but prevents a secrecy claim without verified access controls. The records neither prove exposure nor establish gold/verifier isolation for every historical run. Unlike periodically refreshed evaluations such as LiveBench \citep{white2025livebench}, this static suite does not implement ongoing contamination mitigation. The package reproduces aggregates and selected case extractions, not every artifact judgment or UI trajectory. A complete artifact-level release must pin inputs, verifiers, and historical output hashes and respect source licenses \citep{gebru2021datasheets}. Selective retraction, reconstruction controls, and blind human validation remain future studies, not reported results.

\section{Conclusion}
Reliable Office maintenance requires completing every required update while preserving protected state and native editability. \bench{} makes this contract explicit and connects strict acceptance to live formulas, consistent definitions, retained conditions, and complete delivery. Its diagnostics motivate scope-aware agents; frozen verifier outcomes remain an operational measure requiring independent human calibration.

\section*{Reproducibility statement}
The source package supplies de-identified snapshots, source digests, analysis scripts, a selection ledger, and hash-linked case evidence. The public benchmark is available at \url{https://github.com/Aniriswu/OfficeEditBench}. A separate ten-case evidence bundle includes unchanged inputs and inspected outputs, not all 170 executable tasks. Report-derived tables and current artifact hashes do not establish historical execution replay; that requires complete version-pinned fixtures, verifiers, outputs, and execution manifests.

\section*{Ethics and AI-use statements}
Approvals and personnel are synthetic; public sources retain their attribution and use conditions. Analysis files omit account identifiers and private conversation links. AI assisted drafting, source review, code, artifact extraction, and Figures~\ref{fig:architecture} and~\ref{fig:motivation}; statistical plots use frozen records. AI review is not independent human annotation. Authors remain responsible for claims, permissions, and submission decisions; benchmark results provide no deployment-safety assurance.

\bibliography{references}
\bibliographystyle{iclr2027_conference}
\clearpage
\appendix
\raggedbottom
\section{Evidence ledger and snapshot selection}
\label{app:evidence}
Table~\ref{tab:ledger} specifies the evidence used in each cell of the main results table. WorkBuddy's original score JSON and mechanically generated reports cover all 170 tasks; every composite, strict verdict, and previously available component score agrees with the frozen tables. These records support check-level analysis, not complete execution replay.

\begin{table}[H]
\centering\small
\caption{Evidence available for the September 26 freeze. ``First delivery'' is a recorded selection rule, not a verified universal first-attempt claim.}
\label{tab:ledger}
\begin{tabularx}{\textwidth}{p{.22\textwidth}XX}
\toprule
System / tier & Inspected evidence & Selection and uncertainty\\
\midrule
WorkBuddy / Easy & Original score/check JSON and generated task report; 50 tasks & Final formal UI file cards; retries included. 50 readable outputs in stored checks.\\
WorkBuddy / Medium & Original score/check JSON with $C/S/P$; 70 tasks & Formal delivery for all 70; DNS retries on 055/056; authorized candidate B on 063.\\
WorkBuddy / Hard & Original score/check JSON; 50 tasks; exploratory designation retained & 46 package-valid triples; four missing-submission zeros.\\
Doubao / Easy & Score/check JSON and progress metadata & Fresh conversation; first delivery selected according to metadata.\\
Doubao / Medium & Two progress files: 20 + 50 task scores & Completed 001--070; component means not available in the imported records.\\
Doubao / Hard & Final score/check JSON and historical summary & Final corrected snapshot after reruns and interaction corrections.\\
Codex / Easy & Score/check JSON; model label in run directory & Attempt-selection history not established by this score file.\\
Codex / Medium & Score/check JSON and protocol metadata & Independent projectless task per item; first delivery selected.\\
Codex / Hard & Score/check JSON; model label in run directory & Attempt-selection history not established by this score file.\\
\bottomrule
\end{tabularx}
\end{table}

\paragraph{Retry details.}
WorkBuddy Easy E2-003 reached formal delivery on the third attempt and E2-004 on the second; E2-029--032 were retried after network/UI problems. These attempts are not recoded as first-attempt successes. WorkBuddy Medium excludes initial DNS failures on OBM-055/056 from its final-delivery record. We retain these recorded attempt histories and selection rules rather than reconstructing a common first-attempt rate.

\paragraph{Historical changes.}
The earlier Doubao Medium prefix was 30/63, with mean 77.54; the present full snapshot is 35/70, with mean 78.36. These are different coverage freezes, not repeated trials. Hard Doubao historical means include 41.48, 40.14, 43.35, and 43.52; every stored stage has zero strict passes. Reruns, interaction corrections, and evaluation changes prevent attributing that trajectory to a single capability improvement. The designated final snapshot is reported without pooling history.

\paragraph{Aggregation and missingness.}
All composites are equal-task averages of stored scores, which may already be rounded. WorkBuddy Hard has composite sum 2294.62 over 50 tasks; its $S/P$ means exclude four unavailable components. Codex Hard $S/P$ means include OBW4-045's zeros. Conditioning Codex on its 49 package-valid triples gives $S=44.68$ and $P=81.47$, rather than the all-task 43.78 and 79.84. Doubao has 50 such triples, so its conditional and all-task means coincide. We do not reconstruct composite means by weighting aggregate $S/P$: weights vary by task type, and denominators can differ.

\paragraph{Preservation-count definition.}
The WorkBuddy Medium records contain 22 tasks with $P<100$. Eighteen fail the check named \texttt{preservation}; the remaining failures occur under other check identifiers. Thus a count restricted to that name differs from all preservation-component failures. We use the component-based total of 22. The 33 rejected tasks partition into eight target-perfect preservation failures, eleven preservation-perfect target failures, and fourteen with both components below 100. The raw-record import does not alter the frozen scores.

\subsection{WorkBuddy operator-supplied case analysis}
\label{app:workbuddy-cases}
WorkBuddy's evidence comprises original Easy verification JSON, Medium score JSON, Hard final-score JSON, and three mechanically generated per-task reports. The records contain 200 Easy, 464 Medium, and 1,987 Hard atomic checks; source digests and per-task reconciliation are included in the supplement. The case discussion separates what a record establishes from the operator's explanation of the editing mechanism. The archived Hard run retains its \emph{exploratory} designation. These score records establish check outcomes, not matched execution conditions.

OBM-032 passes all stored semantic and preservation checks. OBM-034 passes recalculation and native chart-data checks but fails \texttt{editable\_formulas} and \texttt{formula\_reacts\_to\_source}; the operator identifies formula-to-constant replacement as the cause, which the logs alone do not trace. OBM-058's record contains the actual on-time classifications at paragraph 4 and table 0, cell (1,2), contrary to the required delayed classification. OBM-062 records actual prose and table text that omit the required invoice condition. These expected/actual strings support the specific textual differences without a new native rendering inspection or a mechanism-frequency estimate.

The Easy E2-013 record fails target-format checks at \texttt{Rates!B2} and \texttt{Summary!B4}, as well as non-target preservation at \texttt{Purchases!A1:E1}; the operator reports that the requested values were changed correctly, but those values are not stored in the check details. E2-030 fails target-text and non-target text-format checks; the paraphrasing account comes from the operator. All five DOCX cases fail non-target checks; E2-046, 047, and 049 also fail target checks. Their basic structure checks pass and no actual style IDs are recorded. The reported renumbering explanation is therefore not an independently established cause. OBW4-050 and 002 contain failed note predicates with nonempty actual notes, so failure is not automatically an absent note. OBW4-006 records a missing submission, with $S/P$ unscored rather than zero.

The companion atomic-record JSON files preserve check identifiers, weights, outcomes, and messages. De-identified provenance records source hashes and exact score reconciliation; an operator addendum retains contextual mechanism descriptions. Original aggregate scores and retry records remain unchanged.

\subsection{Artifact-level case inspection}
\label{app:recovered}
WorkBuddy case analysis uses archived atomic checks; direct artifact inspection covers the Codex cases below, with inspected-file hashes supplied as evidence rather than proof of historical execution replay.

The Easy inspection covers E2-001, 002, 003, 004, 007, 008, 010, and 011. Their materialized input hashes match task metadata. Both library extraction and \texttt{xl/workbook.xml} identify \texttt{Lists} as hidden in the input and visible in the output; the implemented signature's sheet order, freeze panes, merged ranges, tables, chart/image counts, and hidden rows/columns otherwise match. This is a reproducible state difference, not a claim that the workbook is unusable or that hiding the sheet would repair every possible problem.

The Medium inspection maps slide relationships to native chart, embedded workbook, and notes parts rather than inferring values from screenshots. OBM-037's January chart value is 20.075477 million MWh in both cache and embedded workbook, exactly matching 20,075,477 MWh; February remains 17.7412. The reference rounds January to 20.0755, 23 MWh higher. The frozen numeric comparison uses relative tolerance $10^{-8}$ and absolute tolerance $10^{-7}$, rejecting this precision difference; only \texttt{chart\_data\_6} fails in the record. OBM-035's slide-6 notes give numerator 989, denominator 1396, rate 70.8453\%, and explicitly exclude four cancelled flights. The \texttt{notes\_6} rule requires reference numeric substrings and a literal Chinese phrase meaning ``already excluded''; the numbers are present but that phrase is absent. Its added \texttt{Review!A5/B5} cancellation row is also marked non-target by the checker, although the request asks to list cancelled flights separately; raw \texttt{Flights} XML is unchanged. These observations flag equivalence and authorization questions, not complete human acceptance. Other OBM-035 scope failures remain unadjudicated.

The included case JSON files retain instructions, exact extracted values, file/part locators, frozen verdicts, and source hashes. Extraction is purposive and performed programmatically with AI assistance; it is not blind human annotation, a native rendering test, or an error-rate estimate. Original files, verifiers, and scores were not modified.

\section{Dataset composition and intended use}
\label{app:dataset}
The active set uses Easy v2, not the retired Easy tasks. Easy targets are deliberately narrow: XLSX dates, values, assumptions and formulas; PPTX cover/body copy, color, and a native table; DOCX table headers, titles, and action lines. Bilingual coverage is a property of the set, not a paired translation experiment. Language, format, and specific requests co-vary; language-slice scores do not identify a language capability effect.

\begin{table}[H]
\centering\small
\caption{Hard fixture sizes reported by the repository. Pages describe construction metadata, not renderer-invariant pagination; rows and slides describe shared fixtures, not independent samples.}
\begin{tabular}{lrrrr}
\toprule
Scenario & Task IDs & Spreadsheet rows & Slides & DOCX pages\\
\midrule
Public-company reporting & 13 & 160,000 & 112 & 86\\
Retail operations & 13 & 220,000 & 128 & 92\\
Capital/procurement & 12 & 220,000 & 118 & 90\\
Environmental compliance & 12 & 220,000 & 142 & 104\\
\bottomrule
\end{tabular}
\end{table}

\paragraph{Workflow profiles.}
Eleven Hard profiles have four tasks each: full closure, restatement, hold, policy update, interactive clarification, independent edits, dependent cascade, partial approval, late amendment, evidence repair, and scoped clarification. Three profiles have two tasks each: formula-chain release, split release, and duplicate-label correction. The four interactive and four scoped-clarification tasks account for the eight required-clarification instances. Profiles vary authorization and dependency patterns within shared sources; they do not eliminate template dependence.

\paragraph{Source and release boundaries.}
Hard source documentation identifies SEC Financial Statement Data Sets, UCI Online Retail II, City of Chicago contracts/building-permit data, and EPA ECHO-derived material. Medium cohorts document their own public-data sampling and synthetic narratives. Original source records remain subject to their own terms and attribution requirements. The manuscript analysis package contains de-identified derived records rather than redistributing every original source file. A full benchmark release must preserve source attribution and distinguish synthetic controls from historical business records.

The current suite is suitable for local diagnostics and verifier research, not contamination-resistant ranking or unattended consequential deployment. New evaluation tasks should be held out by source workspace or scenario family, rather than by randomly splitting highly related task IDs. Revisions motivated by observed failures should be disclosed and evaluated uniformly across systems.

\paragraph{Contract size is not an independent difficulty measure.}
The static Hard inventory distinguishes gold operation specifications from records emitted by the verifier (Table~\ref{tab:contract-inventory}). The 50 tasks contain 14--43 gold operations each, of which 9--43 specify target checks and 0--8 explicit preservation checks. Global invariants add checks beyond these specifications. Package-valid Codex/Doubao records store 34--63 checks per task, including 9--43 semantic and 20--28 preservation records. Missing-file sentinels are reported separately rather than interpreted as observed damage. Dependency-depth values are author-declared labels, not measured graph depths; distinct edit sites, dependency edges, and protected-object counts were not enumerated. All Hard strict outcomes are zero at every observed contract length; that fact does not isolate a causal effect of conjunction length or file size.

\begin{table}[t]
\centering\small
\caption{Hard contract specifications and stored check records. Counts are not numbers of edited or protected objects. Source hashes and per-task distributions are provided in \texttt{contract\_inventory.json}.}
\label{tab:contract-inventory}
\begin{tabular}{@{}p{.54\linewidth}p{.39\linewidth}@{}}
\toprule
Quantity & Recorded value \\
\midrule
Target-check specifications / task & 9--43 \\
All gold operation specifications / task & 14--43 \\
Explicit preservation specifications / task & 0--8 \\
Declared depth labels (tasks) & 4 (16); 5 (28); 7 (6) \\
Dependency edges / distinct edit sites & Not enumerated \\
Protected-object count & Not enumerated \\
\midrule
Codex, missing-file (1) & 49 total; S 20; P 26 \\
Codex, package-valid (49) & 34--63 total; S 9--43; P 20--28 \\
Doubao, package-valid (50) & 34--63 total; S 9--43; P 20--28 \\
\bottomrule
\end{tabular}
\par\vspace{2pt}\begin{minipage}{.96\linewidth}\footnotesize S/P count stored semantic/preservation records, excluding file sentinels. Codex's missing-file row contains three additional file-failure sentinels; its placeholder failures are not independently tested artifact predicates. Interaction summaries are stored, but their constituent check arrays are not. Public sources ground selected file contexts; requests are benchmark-authored, with no documented real-user sampling.\end{minipage}
\end{table}

\section{Cohort and preservation diagnostics}
\label{app:cohorts}
\begin{figure}[H]
\centering
\includegraphics[width=\textwidth]{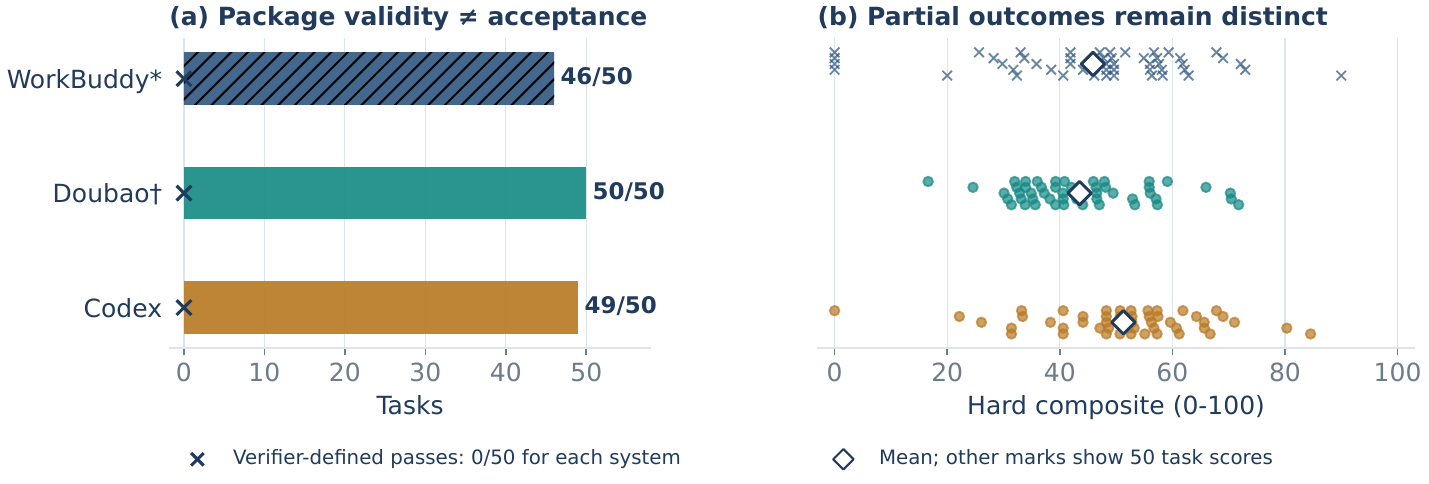}
\caption{\textbf{Delivery versus verifier-defined acceptance on Hard.} (a) Package-valid triples versus zero strict passes. (b) All 50 stored composites per system, including missing-output zeros; diamonds denote means and point offsets aid readability. *WorkBuddy is the exploratory snapshot, corroborated by operator-supplied atomic records; $\dagger$Doubao is the corrected snapshot. Package validity does not certify native-application editability.}
\label{fig:gap}
\end{figure}

\begin{table}[H]
\centering\small
\caption{Easy v2 by file format. Each pair is strict passes and mean score. All three systems have archived check-level evidence. Equal means can conceal different pass counts.}
\label{tab:easy}
\begin{tabular}{lrrrrrr}
\toprule
 & \multicolumn{2}{c}{WorkBuddy} & \multicolumn{2}{c}{Doubao} & \multicolumn{2}{c}{Codex}\\
Format & Pass & Mean & Pass & Mean & Pass & Mean\\
\midrule
XLSX & 19/20 & 96.00 & 19/20 & 99.50 & 12/20 & 96.00 \\
PPTX & 19/20 & 96.00 & 1/20 & 14.50 & 20/20 & 100.00 \\
DOCX & 5/10 & 74.00 & 9/10 & 90.00 & 10/10 & 100.00 \\
 
\bottomrule
\end{tabular}
\end{table}

\begin{table}[H]
\centering\small
\caption{Medium cohort results: strict passes and mean composites. All cohorts are complete; WorkBuddy values match its atomic records. Cohorts do not isolate a single causal difficulty variable.}
\begin{tabular}{lrrrrrr}
\toprule
 & \multicolumn{2}{c}{WorkBuddy} & \multicolumn{2}{c}{Doubao} & \multicolumn{2}{c}{Codex}\\
Task IDs & Pass & Mean & Pass & Mean & Pass & Mean\\
\midrule
001--010 & 8/10 & 97.00 & 8/10 & 98.48 & 10/10 & 100.00 \\
011--030 & 11/20 & 80.40 & 11/20 & 82.31 & 13/20 & 88.53 \\
031--050 & 13/20 & 94.84 & 8/20 & 78.08 & 11/20 & 94.29 \\
051--070 & 5/20 & 64.75 & 8/20 & 64.63 & 10/20 & 78.56 \\
 
\bottomrule
\end{tabular}
\end{table}

WorkBuddy's $S=100$ strict failures are OBM-001, 002, 011, 013, 028, 057, 066, and 070; Codex's are OBM-016, 029, 041, 047, and 050. WorkBuddy has 45 tasks with $S=100$ and 48 with $P=100$; Codex has 49 and 61. These are verifier component counts, not independent human judgments. Comparable Doubao component counts are unavailable in the imported Medium progress files.

\begin{figure}[H]
\centering
\includegraphics[width=\textwidth]{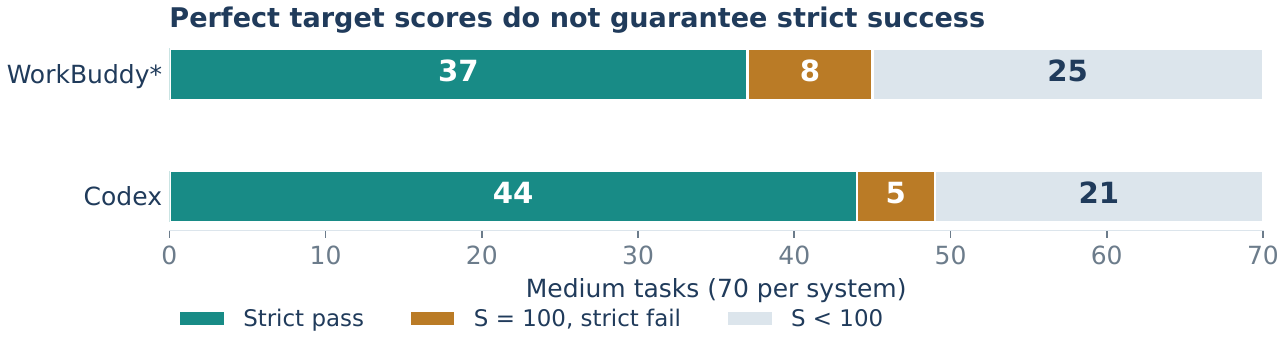}
\caption{\textbf{Target-perfect failures in Medium.} Each bar partitions all 70 tasks into strict passes, perfect semantic scores that still fail strict success, and remaining outcomes. WorkBuddy counts match the supplied atomic records. Doubao is omitted because comparable component scores were not imported. These are verifier outcomes, not human acceptability.}
\label{fig:targetgap}
\end{figure}

\begin{table}[H]
\centering\small
\caption{Non-exclusive Hard preservation failures \emph{among package-valid triples}. Counts use stored predicate verdicts; WorkBuddy's atomic records corroborate its report. A predicate failure is not a count of damaged objects.}
\begin{tabular}{lrrr}
\toprule
Failed predicate & WorkBuddy ($n=46$) & Doubao ($n=50$) & Codex ($n=49$)\\
\midrule
Non-target document text & 34 & 45 & 49\\
Non-target spreadsheet cells/formulas & 33 & 44 & 35\\
Non-target presentation notes & 25 & 23 & 18\\
Non-target presentation text & 20 & 26 & 30\\
\bottomrule
\end{tabular}
\end{table}

The Codex all-task counts are one higher in each row because the missing-output task fails these checks too. We exclude that task here to avoid describing missing files as observed collateral edits. Table~\ref{tab:results} nevertheless retains its zero composite and component scores in the all-task means.

\subsection{Shared-source composition and sensitivity}
\label{app:source-sensitivity}
Table~\ref{tab:hard_sources} groups Hard outcomes using each task's recorded source scenario, not contiguous ID ranges. Table~\ref{tab:hard_source_sensitivity} removes one complete source group at a time. The retained composite means range from 43.58--49.05 for WorkBuddy, 42.40--44.06 for Doubao, and 50.54--52.12 for Codex. These are descriptive composition effects, not uncertainty intervals or generalization results. A held-out source experiment would require unseen source families and new executions. The companion JSON also retains each group's score distribution and available interaction denominators.

\begin{table}[H]
\centering\small
\caption{Hard results grouped by shared source scenario. $D$ is the number of package-valid triples; $n$ counts all tasks. Composite $C$ includes all $n$ stored outcomes. $S_D/P_D$ condition on the same $D$ deliveries (their component denominator is $D$, not $n$).}
\label{tab:hard_sources}
\begin{tabular}{llrrrrr}
\toprule
Source & System & $D/n$ & Strict/$n$ & Mean $C$ & $S_D$ & $P_D$\\
\midrule
Public-company & WorkBuddy & 13/13 & 0/13 & 52.48 & 44.37 & 83.93 \\
 & Doubao & 13/13 & 0/13 & 46.70 & 39.28 & 71.61 \\
 & Codex & 12/13 & 0/13 & 51.10 & 48.83 & 81.96 \\
\midrule
Retail & WorkBuddy & 10/13 & 0/13 & 36.91 & 38.36 & 84.40 \\
 & Doubao & 13/13 & 0/13 & 41.97 & 32.91 & 73.27 \\
 & Codex & 13/13 & 0/13 & 53.44 & 45.42 & 82.21 \\
\midrule
Capital/procurement & WorkBuddy & 12/12 & 0/12 & 46.93 & 37.20 & 82.31 \\
 & Doubao & 12/12 & 0/12 & 42.04 & 32.35 & 76.32 \\
 & Codex & 12/12 & 0/12 & 51.80 & 43.78 & 81.29 \\
\midrule
Environmental & WorkBuddy & 11/12 & 0/12 & 47.45 & 44.82 & 80.34 \\
 & Doubao & 12/12 & 0/12 & 43.23 & 35.74 & 71.61 \\
 & Codex & 12/12 & 0/12 & 48.68 & 40.61 & 80.34 \\
\bottomrule
\end{tabular}
\par\smallskip
\begin{minipage}{\textwidth}\footnotesize WorkBuddy is report-derived; Doubao uses its corrected final snapshot. Package validity is not native-application usability. Component means exclude non-deliveries in this table only; the main table retains its stated all-available denominators. Sources are shared groups, not independent tasks or matched difficulty conditions.\end{minipage}
\end{table}

\begin{table}[H]
\centering\small
\caption{Descriptive leave-one-source-scenario-out sensitivity of all-task Hard composite means. A row removes the complete named source group and reaggregates retained records; it is not a held-out generalization experiment or a confidence interval.}
\label{tab:hard_source_sensitivity}
\begin{tabular}{lrrrr}
\toprule
Omitted source & Retained $n$ & WorkBuddy & Doubao & Codex\\
\midrule
None (full snapshot) & 50 & 45.89 & 43.52 & 51.29 \\
Public-company & 37 & 43.58 & 42.40 & 51.36 \\
Retail & 37 & 49.05 & 44.06 & 50.54 \\
Capital/procurement & 38 & 45.57 & 43.99 & 51.14 \\
Environmental & 38 & 45.40 & 43.61 & 52.12 \\
\bottomrule
\end{tabular}
\par\smallskip
\begin{minipage}{\textwidth}\footnotesize Every row weights retained tasks equally and includes stored non-delivery zeros. No new run, reweighting by scenario, or score revision is performed. The zero strict-pass result persists after each omission, but this is algebraically expected from an all-zero full snapshot and is not independent robustness evidence.\end{minipage}
\end{table}

\section{Measurement validity and confirmatory protocol}
\label{app:measurement}
\begin{figure}[H]
\centering
\includegraphics[width=\textwidth]{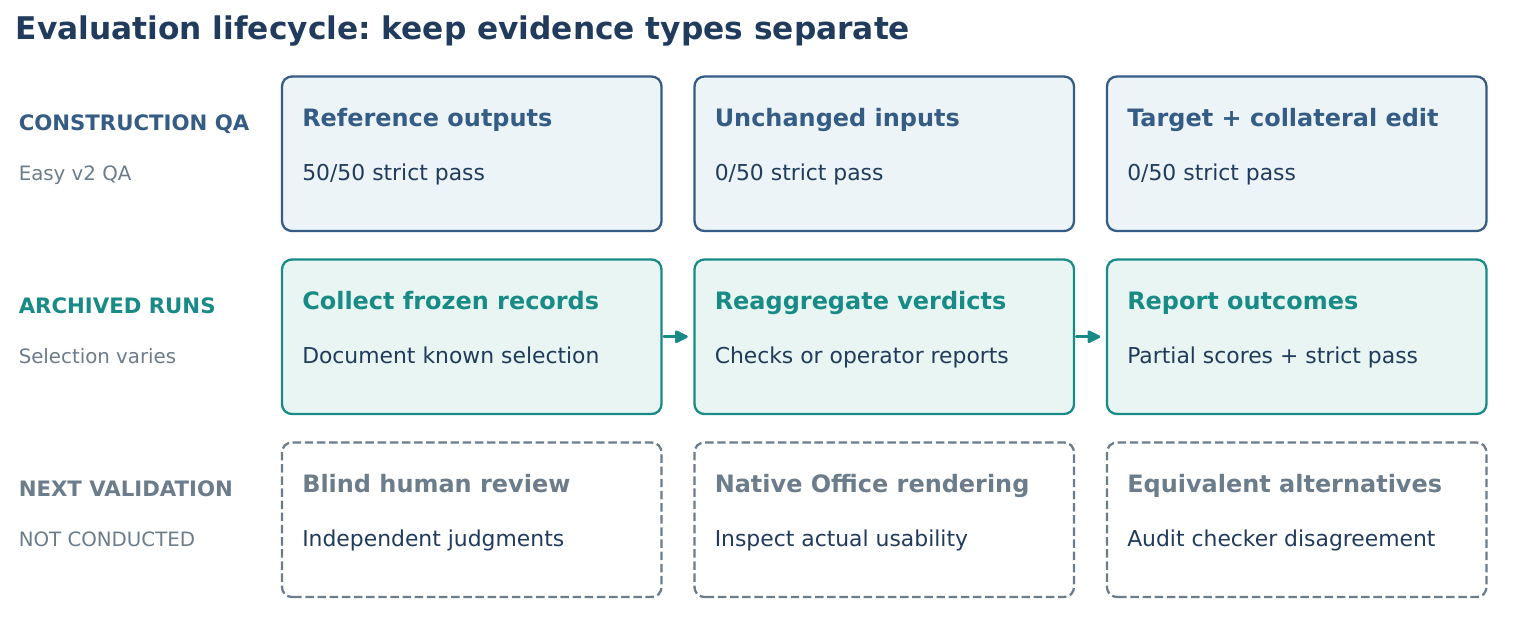}
\caption{\textbf{Evaluation evidence and its limits.} The top row contains three separate Easy v2 construction-QA probes, not an agent execution sequence. The middle row is the archived-record analysis workflow; atomic checks or operator reports are used according to availability. Dashed cards mark proposed independent validation that has not been conducted in this study. Construction QA is not a human baseline.}
\label{fig:lifecycle}
\end{figure}

Figure~\ref{fig:lifecycle} separates construction probes, archived-record analysis, and the independent validation still needed. Atomic logs support the middle stage; they do not substitute for independent validation.

\begin{table}[H]
\centering\small
\caption{What a check can establish, and what requires additional evidence.}
\begin{tabularx}{\textwidth}{p{.18\textwidth}XX}
\toprule
Check family & Measured property & Not established by that check alone\\
\midrule
Readability/package & Expected file is parseable or ZIP-valid under the checker & Native Office opening, recalculation, visual fidelity, accessibility\\
Target & Specified value/type, text, formula, or native object meets a predicate & Every business claim is true; all equivalent answers are accepted\\
Preservation & Selected non-target state matches the starting artifact & Severity of a mismatch; byte or pixel identity of the full file\\
Structure & Selected sheets, objects, paragraph/table structure or identifiers match & A failed structural signature means the output is unusable\\
Interaction & Required concepts and fixed reply satisfy the transcript check & Full pre-reply behavior, historical isolation, universal first attempt\\
\bottomrule
\end{tabularx}
\end{table}

\paragraph{Construction QA is not a human baseline.}
The Easy QA record additionally reports rendering ten DOCX reference outputs and opening two PPTX references in native PowerPoint. Those are limited checks of constructed references, not comprehensive inspection of the 150 Easy agent outputs. The Hard audit's reference generation shares gold assumptions. Its positive/negative regressions test the implemented predicates but cannot estimate their false-rejection rate on all legitimate alternative edits.

\begin{figure}[H]
\centering
\includegraphics[width=\textwidth]{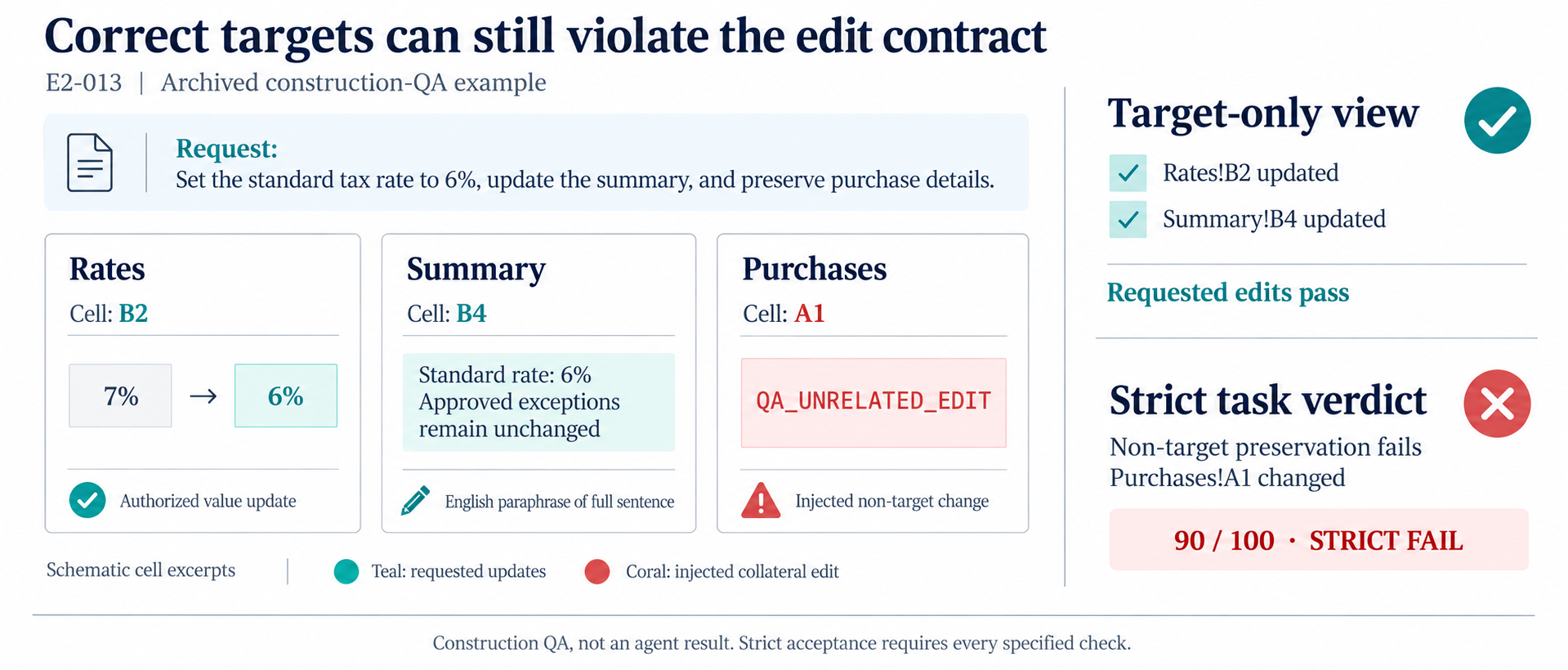}
\caption{\textbf{Construction probe, not an agent failure.} In archived E2-013 QA, both requested updates pass, but an injected overwrite of \texttt{Purchases!A1} fails non-target preservation: 90/100, strict failure. Cells are schematic; the original Chinese Summary sentence is paraphrased in English.}
\label{fig:motivation}
\end{figure}

\paragraph{Recommended independent validation (not conducted).}
Preregister 72 task--system records, each inspected as its entire required artifact set. In each Easy/Medium system cell, randomly sample four automatically accepted and four rejected records; in each Hard system cell, sample two records per source scenario. This yields eight records per system--tier cell without assuming disjoint failure families. Publish stratum sizes, inclusion probabilities, and the random seed. Failure families are subsequent diagnostic slices. Add all eight Codex Easy structural disagreements and five WorkBuddy Easy style-ID cases as a separately reported challenge set. Overlaps retain their probability-sample membership and weight; purposive additions do not enter unweighted population estimates.

Distinguish recorded non-delivery from artifacts that were delivered but cannot now be recovered. Retain the sampled records, reasons for non-response, and actual auditable denominators; weights alone cannot correct archive-availability bias. Feasibility and the population to which results apply depend on recovering actual files, particularly for report-derived records. Hard has no automatically accepted outputs: human rejection conditional on automatic acceptance is undefined. Zero automatic acceptance among human-rejected Hard outputs would be degenerate for this frozen corpus, not evidence that the verifier detects all harmful edits.

Give two qualified reviewers the task, starting files, visible evidence, and delivered artifacts, hiding product identities and automatic verdicts as far as artifact contents permit. Independently label target correctness, authorization, protected-state integrity, native opening/recalculation, and continued editability, recording application/version and supporting locations. Calibrate the rubric on separate examples; retain independent labels before adjudication. Report agreement and criterion-level confusion matrices with explicit denominators; separate verifier rejection among human-acceptable outputs from acceptance among human-rejected outputs. Use sampling weights for population summaries and disclose shared-source clustering and small-sample uncertainty. Review both accepted and rejected outputs: inspecting only failures cannot measure false acceptance. This is a proposed study, not an annotation result.

\paragraph{Complementary evaluator challenge sets (not conducted).}
One set should contain functionally equivalent edits with different representations: legal paraphrases, equivalent formulas, and harmless identifier renumbering. A second should retain selected checked values while violating the task's substantive intent, rendering, or native editability. The first probes excessive rejection; the second probes incomplete coverage. Freeze the original verdicts, report independently established equivalence or harm, and evaluate any revised checker as a new version across all affected systems. Neither challenge set is a substitute for sampling actual agent outputs.

\paragraph{Minimum execution manifest.}
An auditable rerun should record task/input/evidence hashes, exact instruction, product and model configuration, session isolation, permitted tools, start/stop time, action budget, first reply, fixed clarification reply if applicable, all retry reasons, selected output hashes, verifier version/hash, and per-check verdicts. Preserve first attempts and recovery attempts as distinct records. Define infrastructure exclusions before observing success. Historical source-file digests cannot substitute for these runtime fields.

\paragraph{Future extensions.}
Selective retraction would test removing one approved change while retaining another and recomputing dependencies. Reconstruction would test building the same target from corrected inputs under a matched budget. Neither condition appears in the empirical tables here. An extension would need independently derived target states and separate forward, reference-start, and chained outcomes; a worked semantic example alone would not establish agent performance.

\clearpage
\section{Complete Easy v2 outcomes}
Each row gives stored score and verifier strict pass (Y/N). IDs have prefix E2. WorkBuddy's original check JSON corroborates every previously frozen entry. The two horizontal panels are consecutive task ranges, not separate experiments.

\begin{table}[H]
\centering\footnotesize
\setlength{\tabcolsep}{4pt}
\caption{All 50 Easy v2 tasks, three systems (150 outcomes).}
\begin{tabular}{rrcrcrc@{\hspace{12pt}}rrcrcrc}
\toprule
 & \multicolumn{2}{c}{WorkBuddy} & \multicolumn{2}{c}{Doubao} & \multicolumn{2}{c}{Codex} & & \multicolumn{2}{c}{WorkBuddy} & \multicolumn{2}{c}{Doubao} & \multicolumn{2}{c}{Codex}\\
ID & Score & Pass & Score & Pass & Score & Pass & ID & Score & Pass & Score & Pass & Score & Pass\\
\midrule
001 & 100.00 & Y & 100.00 & Y & 90.00 & N & 026 & 100.00 & Y & 10.00 & N & 100.00 & Y \\
002 & 100.00 & Y & 100.00 & Y & 90.00 & N & 027 & 100.00 & Y & 100.00 & Y & 100.00 & Y \\
003 & 100.00 & Y & 100.00 & Y & 90.00 & N & 028 & 100.00 & Y & 10.00 & N & 100.00 & Y \\
004 & 100.00 & Y & 100.00 & Y & 90.00 & N & 029 & 100.00 & Y & 10.00 & N & 100.00 & Y \\
005 & 100.00 & Y & 100.00 & Y & 100.00 & Y & 030 & 20.00 & N & 10.00 & N & 100.00 & Y \\
006 & 100.00 & Y & 100.00 & Y & 100.00 & Y & 031 & 100.00 & Y & 10.00 & N & 100.00 & Y \\
007 & 100.00 & Y & 100.00 & Y & 90.00 & N & 032 & 100.00 & Y & 10.00 & N & 100.00 & Y \\
008 & 100.00 & Y & 100.00 & Y & 90.00 & N & 033 & 100.00 & Y & 10.00 & N & 100.00 & Y \\
009 & 100.00 & Y & 100.00 & Y & 100.00 & Y & 034 & 100.00 & Y & 10.00 & N & 100.00 & Y \\
010 & 100.00 & Y & 100.00 & Y & 90.00 & N & 035 & 100.00 & Y & 10.00 & N & 100.00 & Y \\
011 & 100.00 & Y & 100.00 & Y & 90.00 & N & 036 & 100.00 & Y & 10.00 & N & 100.00 & Y \\
012 & 100.00 & Y & 100.00 & Y & 100.00 & Y & 037 & 100.00 & Y & 10.00 & N & 100.00 & Y \\
013 & 20.00 & N & 100.00 & Y & 100.00 & Y & 038 & 100.00 & Y & 10.00 & N & 100.00 & Y \\
014 & 100.00 & Y & 100.00 & Y & 100.00 & Y & 039 & 100.00 & Y & 10.00 & N & 100.00 & Y \\
015 & 100.00 & Y & 100.00 & Y & 100.00 & Y & 040 & 100.00 & Y & 10.00 & N & 100.00 & Y \\
016 & 100.00 & Y & 90.00 & N & 100.00 & Y & 041 & 100.00 & Y & 100.00 & Y & 100.00 & Y \\
017 & 100.00 & Y & 100.00 & Y & 100.00 & Y & 042 & 90.00 & N & 0.00 & N & 100.00 & Y \\
018 & 100.00 & Y & 100.00 & Y & 100.00 & Y & 043 & 100.00 & Y & 100.00 & Y & 100.00 & Y \\
019 & 100.00 & Y & 100.00 & Y & 100.00 & Y & 044 & 100.00 & Y & 100.00 & Y & 100.00 & Y \\
020 & 100.00 & Y & 100.00 & Y & 100.00 & Y & 045 & 90.00 & N & 100.00 & Y & 100.00 & Y \\
021 & 100.00 & Y & 10.00 & N & 100.00 & Y & 046 & 20.00 & N & 100.00 & Y & 100.00 & Y \\
022 & 100.00 & Y & 10.00 & N & 100.00 & Y & 047 & 20.00 & N & 100.00 & Y & 100.00 & Y \\
023 & 100.00 & Y & 10.00 & N & 100.00 & Y & 048 & 100.00 & Y & 100.00 & Y & 100.00 & Y \\
024 & 100.00 & Y & 10.00 & N & 100.00 & Y & 049 & 20.00 & N & 100.00 & Y & 100.00 & Y \\
025 & 100.00 & Y & 10.00 & N & 100.00 & Y & 050 & 100.00 & Y & 100.00 & Y & 100.00 & Y \\
 
\bottomrule
\end{tabular}
\end{table}

The eight Codex structural failures are E2-001, 002, 003, 004, 007, 008, 010, and 011. All other Easy Codex checks pass. WorkBuddy's reported style-ID cases are E2-042, 045, 046, 047, and 049. The five style-ID cases are not reclassified as successes without an independent equivalence judgment.

\clearpage
\section{Complete Medium outcomes}
IDs have prefix OBM. Every system has a stored result for each of the 70 tasks. Scores preserve the selected snapshot's delivery-contract failures and retry policy; they are not reconstructed first-attempt scores.

\begin{table}[H]
\centering\footnotesize
\setlength{\tabcolsep}{4pt}
\caption{All 70 Medium tasks, three systems (210 outcomes).}
\begin{tabular}{rrcrcrc@{\hspace{12pt}}rrcrcrc}
\toprule
 & \multicolumn{2}{c}{WorkBuddy} & \multicolumn{2}{c}{Doubao} & \multicolumn{2}{c}{Codex} & & \multicolumn{2}{c}{WorkBuddy} & \multicolumn{2}{c}{Doubao} & \multicolumn{2}{c}{Codex}\\
ID & Score & Pass & Score & Pass & Score & Pass & ID & Score & Pass & Score & Pass & Score & Pass\\
\midrule
001 & 85.00 & N & 100.00 & Y & 100.00 & Y & 036 & 87.50 & N & 100.00 & Y & 100.00 & Y \\
002 & 85.00 & N & 100.00 & Y & 100.00 & Y & 037 & 100.00 & Y & 100.00 & Y & 83.93 & N \\
003 & 100.00 & Y & 100.00 & Y & 100.00 & Y & 038 & 100.00 & Y & 100.00 & Y & 100.00 & Y \\
004 & 100.00 & Y & 100.00 & Y & 100.00 & Y & 039 & 76.56 & N & 35.27 & N & 90.62 & N \\
005 & 100.00 & Y & 100.00 & Y & 100.00 & Y & 040 & 100.00 & Y & 100.00 & Y & 100.00 & Y \\
006 & 100.00 & Y & 89.29 & N & 100.00 & Y & 041 & 100.00 & Y & 57.14 & N & 92.86 & N \\
007 & 100.00 & Y & 95.46 & N & 100.00 & Y & 042 & 100.00 & Y & 51.79 & N & 100.00 & Y \\
008 & 100.00 & Y & 100.00 & Y & 100.00 & Y & 043 & 89.29 & N & 62.50 & N & 89.29 & N \\
009 & 100.00 & Y & 100.00 & Y & 100.00 & Y & 044 & 100.00 & Y & 51.79 & N & 100.00 & Y \\
010 & 100.00 & Y & 100.00 & Y & 100.00 & Y & 045 & 100.00 & Y & 58.40 & N & 100.00 & Y \\
011 & 75.00 & N & 100.00 & Y & 100.00 & Y & 046 & 100.00 & Y & 62.50 & N & 100.00 & Y \\
012 & 100.00 & Y & 75.00 & N & 100.00 & Y & 047 & 100.00 & Y & 91.67 & N & 91.67 & N \\
013 & 75.00 & N & 100.00 & Y & 100.00 & Y & 048 & 100.00 & Y & 100.00 & Y & 81.25 & N \\
014 & 100.00 & Y & 100.00 & Y & 100.00 & Y & 049 & 92.86 & N & 100.00 & Y & 85.71 & N \\
015 & 25.00 & N & 50.00 & N & 75.00 & N & 050 & 85.71 & N & 53.57 & N & 92.86 & N \\
016 & 100.00 & Y & 100.00 & Y & 75.00 & N & 051 & 100.00 & Y & 0.00 & N & 100.00 & Y \\
017 & 0.00 & N & 25.00 & N & 0.00 & N & 052 & 81.25 & N & 75.00 & N & 100.00 & Y \\
018 & 100.00 & Y & 100.00 & Y & 100.00 & Y & 053 & 81.25 & N & 100.00 & Y & 100.00 & Y \\
019 & 81.25 & N & 81.25 & N & 81.25 & N & 054 & 50.00 & N & 50.00 & N & 75.00 & N \\
020 & 100.00 & Y & 100.00 & Y & 100.00 & Y & 055 & 56.25 & N & 81.25 & N & 81.25 & N \\
021 & 100.00 & Y & 0.00 & N & 100.00 & Y & 056 & 50.00 & N & 100.00 & Y & 50.00 & N \\
022 & 100.00 & Y & 100.00 & Y & 100.00 & Y & 057 & 75.00 & N & 100.00 & Y & 100.00 & Y \\
023 & 100.00 & Y & 100.00 & Y & 100.00 & Y & 058 & 62.50 & N & 62.50 & N & 81.25 & N \\
024 & 89.29 & N & 100.00 & Y & 89.29 & N & 059 & 50.00 & N & 0.00 & N & 75.00 & N \\
025 & 37.50 & N & 100.00 & Y & 100.00 & Y & 060 & 100.00 & Y & 0.00 & N & 100.00 & Y \\
026 & 50.00 & N & 75.00 & N & 75.00 & N & 061 & 18.75 & N & 43.75 & N & 43.75 & N \\
027 & 100.00 & Y & 100.00 & Y & 100.00 & Y & 062 & 0.00 & N & 25.00 & N & 25.00 & N \\
028 & 75.00 & N & 75.00 & N & 100.00 & Y & 063 & 45.00 & N & 55.00 & N & 40.00 & N \\
029 & 100.00 & Y & 85.00 & N & 75.00 & N & 064 & 100.00 & Y & 100.00 & Y & 100.00 & Y \\
030 & 100.00 & Y & 80.00 & N & 100.00 & Y & 065 & 50.00 & N & 50.00 & N & 50.00 & N \\
031 & 100.00 & Y & 75.00 & N & 100.00 & Y & 066 & 75.00 & N & 100.00 & Y & 100.00 & Y \\
032 & 100.00 & Y & 100.00 & Y & 100.00 & Y & 067 & 100.00 & Y & 100.00 & Y & 100.00 & Y \\
033 & 100.00 & Y & 100.00 & Y & 100.00 & Y & 068 & 100.00 & Y & 100.00 & Y & 100.00 & Y \\
034 & 76.92 & N & 91.35 & N & 100.00 & Y & 069 & 25.00 & N & 50.00 & N & 50.00 & N \\
035 & 87.98 & N & 70.67 & N & 77.56 & N & 070 & 75.00 & N & 100.00 & Y & 100.00 & Y \\
 
\bottomrule
\end{tabular}
\end{table}

\clearpage
\section{Complete Hard outcomes}
IDs have prefix OBW4. All strict verdicts are N. WorkBuddy is the exploratory snapshot with operator-supplied atomic records; Doubao is the final corrected snapshot. The companion JSON files retain semantic, preservation, interaction, and atomic-check fields where recorded; the table shows composites without conflating component denominators.

\begin{table}[H]
\centering\footnotesize
\setlength{\tabcolsep}{4pt}
\caption{All 50 Hard tasks, three systems (150 outcomes).}
\begin{tabular}{rrcrcrc@{\hspace{12pt}}rrcrcrc}
\toprule
 & \multicolumn{2}{c}{WorkBuddy} & \multicolumn{2}{c}{Doubao} & \multicolumn{2}{c}{Codex} & & \multicolumn{2}{c}{WorkBuddy} & \multicolumn{2}{c}{Doubao} & \multicolumn{2}{c}{Codex}\\
ID & Score & Pass & Score & Pass & Score & Pass & ID & Score & Pass & Score & Pass & Score & Pass\\
\midrule
001 & 72.14 & N & 70.29 & N & 66.71 & N & 026 & 72.95 & N & 43.00 & N & 52.09 & N \\
002 & 61.82 & N & 46.55 & N & 56.73 & N & 027 & 47.76 & N & 33.85 & N & 33.21 & N \\
003 & 49.63 & N & 39.25 & N & 48.27 & N & 028 & 28.25 & N & 33.14 & N & 55.11 & N \\
004 & 49.63 & N & 39.25 & N & 64.27 & N & 029 & 40.60 & N & 35.21 & N & 40.60 & N \\
005 & 62.88 & N & 56.06 & N & 61.21 & N & 030 & 47.12 & N & 34.91 & N & 47.12 & N \\
006 & 0.00 & N & 71.77 & N & 71.03 & N & 031 & 33.05 & N & 30.75 & N & 31.41 & N \\
007 & 0.00 & N & 36.00 & N & 60.73 & N & 032 & 0.00 & N & 40.72 & N & 65.65 & N \\
008 & 48.27 & N & 38.24 & N & 48.27 & N & 033 & 44.12 & N & 32.85 & N & 44.12 & N \\
009 & 57.29 & N & 37.22 & N & 57.29 & N & 034 & 29.81 & N & 49.47 & N & 50.73 & N \\
010 & 51.62 & N & 55.89 & N & 51.26 & N & 035 & 41.89 & N & 33.92 & N & 38.34 & N \\
011 & 58.12 & N & 59.11 & N & 65.66 & N & 036 & 25.66 & N & 42.75 & N & 52.83 & N \\
012 & 56.73 & N & 57.09 & N & 59.64 & N & 037 & 32.39 & N & 31.41 & N & 33.38 & N \\
013 & 49.63 & N & 39.25 & N & 48.27 & N & 038 & 69.00 & N & 55.88 & N & 69.00 & N \\
014 & 48.27 & N & 16.61 & N & 56.27 & N & 039 & 38.48 & N & 24.58 & N & 44.12 & N \\
015 & 61.39 & N & 42.07 & N & 50.69 & N & 040 & 0.00 & N & 48.14 & N & 50.73 & N \\
016 & 67.82 & N & 65.97 & N & 67.82 & N & 041 & 41.89 & N & 33.92 & N & 40.60 & N \\
017 & 54.91 & N & 35.64 & N & 57.45 & N & 042 & 49.19 & N & 45.97 & N & 52.83 & N \\
018 & 57.63 & N & 46.58 & N & 48.27 & N & 043 & 33.70 & N & 30.10 & N & 31.41 & N \\
019 & 48.27 & N & 40.61 & N & 57.29 & N & 044 & 62.18 & N & 46.45 & N & 55.88 & N \\
020 & 59.33 & N & 47.92 & N & 55.65 & N & 045 & 56.00 & N & 53.33 & N & 0.00 & N \\
021 & 20.00 & N & 32.33 & N & 48.67 & N & 046 & 56.00 & N & 52.92 & N & 22.15 & N \\
022 & 58.28 & N & 47.02 & N & 61.87 & N & 047 & 35.86 & N & 40.54 & N & 52.66 & N \\
023 & 41.89 & N & 31.98 & N & 40.60 & N & 048 & 46.13 & N & 44.07 & N & 52.66 & N \\
024 & 56.36 & N & 40.85 & N & 53.21 & N & 049 & 48.94 & N & 70.47 & N & 84.55 & N \\
025 & 31.74 & N & 36.74 & N & 26.08 & N & 050 & 90.00 & N & 57.33 & N & 80.33 & N \\
 
\bottomrule
\end{tabular}
\end{table}
\section{Relationship to existing evaluation settings}
\label{app:related}
Table~\ref{tab:related} identifies evaluation units and their relationship to Office maintenance. It is a positioning comparison, not an exhaustive capability inventory: an unlisted property should not be read as absent from a prior benchmark. Published scores on these different tasks are not baselines for \bench{}.

\begin{table}[H]
\centering\small
\caption{Evaluation settings relevant to preservation-aware Office editing. ICLR status refers to the cited formal publication, not an arXiv submission or workshop.}
\label{tab:related}
\begin{tabularx}{\textwidth}{>{\raggedright\arraybackslash}p{.21\textwidth}LL}
\toprule
Benchmark & Evaluation unit & Relationship to the present study\\
\midrule
SWE-bench (ICLR 2024) & Repository patch evaluated with issue-resolution and regression tests & Prior maintenance principle; our protected observations concern native Office state.\\
WebArena (ICLR 2024) & Web task evaluated by functional outcomes & Persistent-environment evaluation; our deliverables are existing files after scoped edits.\\
$\tau$-bench (ICLR 2025) & User/tool episode, database end state, repeated-trial reliability & State and interaction constraints; our fixed clarification checks do not measure the same reliability statistic.\\
MLE-bench (ICLR 2025) & ML competition submission evaluated against held-out targets & Submission-based interface and explicit execution reporting; nominal limits do not replace per-attempt records.\\
ScienceAgentBench (ICLR 2025) & Scientific programming task with automatic and human assessment & Validity evidence beyond automatic outcomes; our construction QA is not independent human validation.\\
OfficeEditBench (this work) & Authorized edit to native Office artifacts with protected state & 170-task suite, decomposed predicates, and a descriptive audit of selected archived outputs.\\
\bottomrule
\end{tabularx}
\end{table}

Task execution, scoring implementation, and measurement validity are separate design questions. The comparison motivates explicit reporting of all three, rather than defining novelty by the number of check families.

\begin{table}[H]
\centering\small
\caption{\textbf{Closest preservation-aware Office evaluations.} Descriptions indicate the reported scope, not an exhaustive feature checklist. Different units and denominators preclude comparing their published scores directly.}
\label{tab:office-related}
\begin{tabularx}{\textwidth}{>{\raggedright\arraybackslash}p{.20\textwidth}LL}
\toprule
Benchmark & Artifact scope and scoring & Preservation and validity evidence\\
\midrule
PPTArena \citep{ofengenden2026pptarena} & Existing PPTX editing; structure and visual judges & Instruction/quality assessment with human alignment; focused on presentations.\\
PPT-Eval \citep{gandhi2026ppteval} & PPTX creation/editing; code/model rubrics and partial credit & Extra-change penalties; meta-evaluation on human-constructed completion attempts.\\
DeckEdit-Bench \citep{kim2026editppt} & Native PPTX editing; page targeting and object preservation & Preservation of out-of-scope objects, distinct from task-level conjunctive acceptance.\\
OmegaUse-OfficeVal \citep{zhou2026omegauseofficeval} & Multi-format Office/PDF deliverables; usability gate then weighted rubrics & Negative criteria for unintended damage; expert--code calibration and a shared execution scaffold.\\
OfficeEditBench & Native XLSX/PPTX/DOCX; scoped updates and protected-state predicates & Conjunctive acceptance plus component diagnostics; construction QA, but no completed independent agent-output adjudication.\\
\bottomrule
\end{tabularx}
\end{table}

Preservation-aware Office evaluation is therefore an overlapping research area, not an unoccupied one. The present contribution lies in the instantiated task contracts and diagnostic suite, rather than a claim to originate file-level evaluation, multi-format office work, or non-target preservation. Independent validity evidence available in some adjacent work remains an important complement to the present check-level analysis.

\end{document}